\documentclass[aps,pre,floats,twocolumn,showpacs,superscriptaddress]{revtex4}
\usepackage[svgnames]{xcolor} % Specify colors by their 'svgnames', for a full list of all colors available see here: http://www.latextemplates.com/svgnames-colors
\usepackage{graphicx,epsfig}
\usepackage{times}
\usepackage{amssymb,amsmath,multirow,rotate,color}
\usepackage{xcolor}
\usepackage[normalem]{ulem}

\usepackage{float}

\begin{document}

\title{Collapse transition in epidemic spreading subject to detection with limited resources}

\author{Santiago Lamata-Otín}
\affiliation{Department of Condensed Matter Physics, University of Zaragoza, 50009 Zaragoza, Spain}
\affiliation{GOTHAM lab, Institute of Biocomputation and Physics of
Complex Systems (BIFI), University of Zaragoza, 50018 Zaragoza, Spain}

\author{Adriana Reyna-Lara}
\affiliation{Instituto Tecnol\'ogico y de Estudios Superiores de Monterrey, 64849 Monterrey, N.L., México}

\author{David Soriano-Pa\~nos}
\affiliation{Institute Gulbenkian of Science (IGC), 2780-156 Oeiras (Portugal)}
\affiliation{GOTHAM lab, Institute of Biocomputation and Physics of
Complex Systems (BIFI), University of Zaragoza, 50018 Zaragoza, Spain}

\author{Vito Latora}
\affiliation{School of Mathematical Sciences, Queen Mary University of London, London E1 4NS, United Kingdom}
\affiliation{Dipartimento di Fisica ed Astronomia, Universit\`a di Catania and INFN, Catania I-95123, Italy}
\affiliation{Complexity Science Hub Vienna, A-1080 Vienna, Austria}

\author{Jes\'us G\'omez-Garde\~nes}
\affiliation{Department of Condensed Matter Physics, University of Zaragoza, 50009 Zaragoza, Spain}
\affiliation{GOTHAM lab, Institute of Biocomputation and Physics of
Complex Systems (BIFI), University of Zaragoza, 50018 Zaragoza, Spain}

%\date{\today}

\begin{abstract}
 Compartmental models are the most widely used framework for modeling infectious diseases. These models  have been continuously refined to  incorporate all the realistic mechanisms that can shape the course of an epidemic outbreak. Building on a compartmental model that accounts for early detection and isolation of infectious individuals through testing, in this article we focus on the viability of detection processes under limited availability of testing resources, and we study how the latter impacts on the detection rate. Our results show that, in addition to the well-known epidemic transition at ${\mathcal{R}}_0=1$, a second transition occurs at ${\mathcal{R}}^*_0>1$ pinpointing the collapse of the detection system and, as a consequence, the switch from a regime of mitigation to a regime in which the pathogen spreads freely. We characterize the epidemic phase diagram of the model as a function of the relevant control parameters: the basic reproduction number, the maximum detection capacity of the system, and the fraction of individuals in shelter. Our analysis thus provides a valuable tool for estimating the detection resources and the level of confinement needed to face epidemic outbreaks.
\end{abstract}

\pacs{89.20.-a, 89.75.Hc, 89.75.Kd}

\maketitle
\section{Introduction}

The COVID-19 pandemic has affected the entire world, causing significant loss of life, economic hardship, and widespread disruption of social and cultural norms. As a result, it is essential to understand how communicable diseases spread and what can be done to mitigate their impact. Mathematical models are particularly useful in this regard~\cite{estrada2020covid,perra2021non}, as they allow researchers to gain valuable insights into the transmission dynamics of infectious diseases \cite{chinazzi20,france}, inform public health policies \cite{Arenas2020,Goldenfeld2020}, and guide efforts to control \cite{Meidan2021,DellaRossa2020} and prevent future outbreaks \cite{Cryptic}.

Compartmental models are the most widely used framework for modeling infectious diseases \cite{rohanibook,anderson1992infectious}. In these models, a population is divided into compartments or states, with transitions between them mediated by different parameters. While simple and widely used, these basic models, such as the Susceptible-Infectious-Recovered model~\cite{SIR27}, have limitations in accounting for complex demographic or social factors that may impact transmission dynamics and the effectiveness of containment policies.

Over the past years, researchers have made significant efforts to overcome the limitations of compartmental models used for modeling infectious diseases. One approach has been to incorporate various elements that make the models more realistic, thus broadening their range of applicability. These refinements cover the use of complex networks to model interactions through which the pathogen can spread \cite{pastor2015epidemic,wang2017unification}, thus allowing to better capture the heterogeneity of the connection between individuals and its impact on disease transmission dynamics. Another refinement has been its combination with diffusion processes \cite{barbosa2018human}, which mimic mobility flows between densely populated areas \cite{hazarie2021interplay}, enabling the development of metapopulation frameworks to analyze the role of travel and movement patterns in the spread of infectious diseases \cite{colizza4,tizzoni2014use,soriano2022modeling}. Additionally, researchers have coupled the spreading dynamics of infectious diseases with behavioral factors that impact the acceptance of interventions \cite{Funk2010,Wang2016}. This refinement acknowledges the importance of human behavior in the success of disease control measures and allows for the exploration of interventions more likely to be adopted by the population~\cite{Bauch2005,Steinegger2020}. 

Following this line of research, recently new compartmental models have been developed to study how early detection and isolation of infectious individuals through testing and the subsequent activation of contact tracing strategies can interrupt the advance of transmission chains \cite{TTT,kojaku2021effectiveness,Bianconi2020Message,serafino2022digital}. In this study, we explore how limited availability of testing resources alter the viability of detection processes and their impact on the ongoing epidemic outbreak. We propose a minimal compartmental model that can simulate the effects of different interventions, such as lockdowns and testing, and derive the epidemic phase diagram analytically.

Our results show that, in addition to the well-known epidemic transition that occurs when the basic reproduction number ${\mathcal{R}}_0$ is ${\mathcal{R}}_0^c=1$, a second transition takes place at ${\mathcal{R}}^*_0>1$, which depends on the maximum detection capacity of the system. When ${\mathcal{R}}_0>{\mathcal{R}}^*_0$, the system moves from a phase where detection can mitigate the epidemic outbreak to a phase where the pathogen spreads freely. By characterizing this transition, we can determine the precise value of ${\mathcal{R}}^*_0$ as a function of both the detection capacity of the system and the fraction of individuals in shelter. Our model provides a valuable tool for estimating the detection resources and confinement needed to face epidemic outbreaks and can be adapted for use in more elaborate models.

\section{Epidemic spreading dynamics}
\label{sec:2}

To simulate the time progression of infections during a single epidemic wave without considering reinfections, we adopt a Susceptible-Infected-Recovered ($SIR$) framework. This compartmental modeling approach divides the population into three epidemiological states: Susceptible ($S$), i.e. individuals who lack prior exposure to the pathogen and, therefore, possess no immunity, Infected ($I$), individuals infected with the pathogen and carrying a sufficient viral load to infect others, and Recovered ($R$), individuals who have overcome the infection and have developed immunity. In addition to the three typical SIR categories, we introduce two further categories: Locked ($L$) and Detected ($D$), so that we can refer to our model as SLIDR. The $L$ category comprises epidemiologically $S$ individuals who are under strict lockdown measures and thus cannot contract the disease. The $D$ group denotes those individuals who were Infected but have been detected through testing, and consequently, are no longer infectious due to their isolation.

\begin{figure}[t!]
\includegraphics[width=0.8\linewidth]{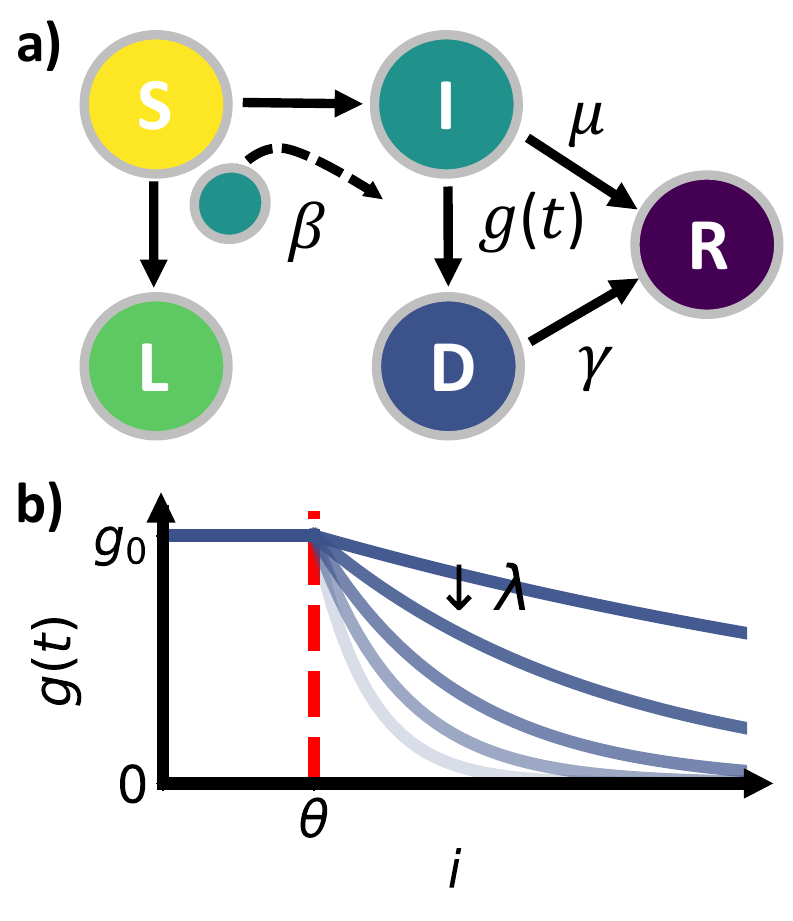}\\% Here is how to import EPS art
\caption{In panel (a) we show the flux diagram of the $SLIDR$ model. The model has five compartments: susceptible ($S$), locked (L), infectious ($I$), detected ($D$) and recovered ($R$). Arrows indicate the possible transitions between different states. In panel (b) we sketch the dependence of the detection rate on the fraction of infected individuals in the population, given by Eq.~(\ref{eq:g_mf}).}
\label{fig1} 
\end{figure}

As typical in compartmental models, each individual of a population can occupy only one state at a given time, and the transition from one state to another is governed by the flow diagram illustrated in Fig.~\ref{fig1}.a. Initially, a fraction $l_0$ of the population transitions from $S$ to $L$, while the remaining individuals in $S$ are susceptible to infection at a rate $\beta$ per contact with an agent in compartment $I$. Infectious individuals transition to either recovered, $R$, at a rate $\mu$, or detected, $D$, at a rate $g(t)$. Note that the detection rate, $g(t)$, is time-dependent, as it is contingent upon the testing capacity, as we will discuss in details below. Finally, detected individuals transition to the recovered state at a rate $\gamma$.

The former transitions allow us to write a set of mean-field equations by considering that each agent is involved in $\langle k\rangle$ contacts per unit time in a population of size $N$. Considering the fractions of the population in each compartment ($s=S/N$, $l=L/N$, $i=I/N$, $d=D/N$, and $r=R/N$) fulfilling $s+l+i+d+r=1$, the differential equations that govern their time evolution read as:
\begin{eqnarray}
%    \frac{ds}{dt}=-(\langle k\rangle-1)\beta si
    \dot s&=&-\left(\langle k\rangle-1\right)\beta si\;,
    \label{eq:s}\\
        \dot l&=&0\;,
    \label{eq:v}\\
    \dot i&=&\left(\langle k\rangle-1\right)\beta si-\left(\mu+g(t)\right) i\;,
    \label{eq:i}\\
        \dot d&=&-\gamma d+gi\;,
    \label{eq:d}\\
    \dot r&=&\mu i+\gamma d\;.
    \label{eq:r}
\end{eqnarray}
Note that, as explained above, Eq.~(\ref{eq:v}) implies that the fraction of population initially set under lockdown remains constant during time ($l(t)=l_0$). 
As mentioned above, the detection rate $g(t)$ is not constant in order to capture the limited nature of testing resources. In particular we assume the time dependence of $g(t)$ shown in Fig.~\ref{fig1}.b, whose mathematical expression reads:
\begin{equation}
    g(t)=
    \begin{cases}
      g_0 & \text{if    }\quad i(t)<\theta\;,\\
      g_0 e^{-\lambda N(i(t)-\theta)} & \text{if   }\quad i(t)>\theta\;.
    \end{cases}       
    \label{eq:g_mf}
\end{equation}
The previous functional form assumes that detection operates in a normal way, i.e. with a constant rate $g_0$, provided that the fraction of infectious agents remains below a specified capacity threshold $\theta$. Under normal conditions, tests are readily available, and the entire detection process, including identification of infectious agents, testing, and processing of results, takes place optimally within an average time period of $1/g_0$. However, when the number of infected individuals $i(t)$ exceeds the capacity threshold $\theta$, we assume that a national health system begins to experience delays, causing a reduction in detection efficiency. Ultimately, when the detection system becomes too slow, infected individuals may recover before being detected, so detection does not alter their infectious period. To model the collapse of the detection system, we introduce an exponential decay of the original rate, $g_0$, towards the detection compartment according to the difference between the availability threshold $\theta$ and the demand $i(t)$. The exponential decay is regulated by a tunable decay rate $\lambda$ times the size of the population $N$. 

\begin{figure*}[t!]
\includegraphics[width=1\linewidth]{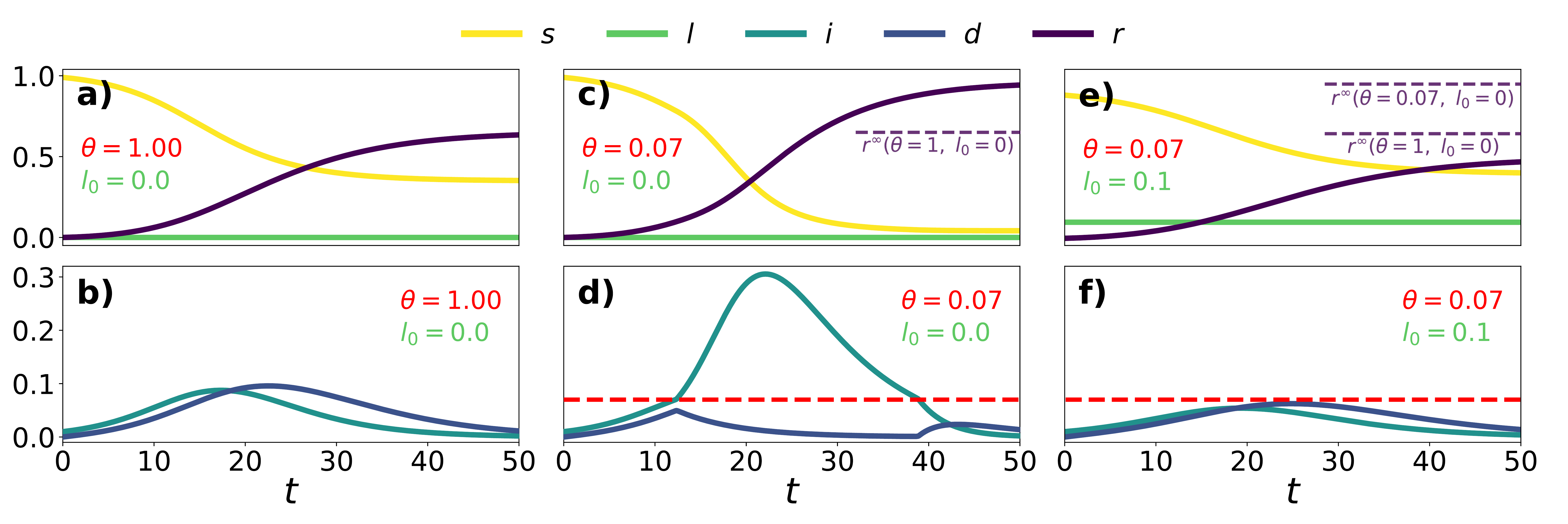}% Here is how to import EPS art
\caption{In panels (a)-(b) we show the temporal evolution of all the compartments of the $SLIDR$ model in absence of lockdown ($l_0=0$), with unlimited resources ($\theta=1$), an average connectivity $\langle k \rangle=5$, a infectivity $\beta=0.137$, a recovery rate $\mu=1/7$, a transition rate from $D$ to $R$ regulated by $\gamma=3/20$, and assuming a baseline detection rate of $g_0=0.2$. In panels (c)-(d) we appreciate how considering limited resources, $\theta=0.07$ (the rest of the parameters are identical to panels (a)-(b)), gives rise to the acceleration of the infected cases growth once the capacity threshold is reached. This acceleration depends on the decay rate $\lambda=1$ and on the population size $N=700$. In panels (e)-(f) we show how collapse can be avoided when a fraction of population is under lockdown ($l_0=0.1$). $r^{\infty}$ reference curves display how lockdown reduces the attack rate of the diseases even below the one corresponding to unlimited resources (panel (a)). The rest of the parameters are identical to panels (c)-(d) ($\theta=0.07$, $\langle k \rangle=5$, $\beta=0.137$, $\mu=1/7$, $\gamma=3/20$, $g_0=0.2$). Note that in all panels the basic reproduction number is fixed to $R_0=1.6$ and time is measured in arbitrary units.}
\label{fig2} 
\end{figure*}

In Fig.~\ref{fig2}, we study how the interplay between the implementation of lockdown policies and the existence of limited testing resources affects epidemic trajectories. To set a reference, we represent an epidemic trajectory in Fig.~\ref{fig2}.a-b with a baseline detection rate $g_0 =0.2$, unlimited resources, i.e. $\theta=1$, and with no lockdown policies at a play, i.e. $l_0=0$. First, we explore in Fig.~\ref{fig2}.c-d the impact of limited resources by setting the capacity threshold to $\theta=0.07$ and the decay rate $\lambda=1$. The selection of the capacity threshold value draws inspiration from the typical testing ratios observed during the most challenging weeks of the COVID-19 epidemic in Europe~\cite{COVIDdata}. The time evolution of each compartment clearly shows that, once the infectious population reaches the value $\theta$ noticed by the horizontal dashed red line in Fig.~\ref{fig2}.d , the fraction of detected agents decays and, as a consequence, the increase of the infectious population speeds up.
As a consequence, a much larger attack rate $r^{\infty}=\lim_{t\rightarrow\infty}r(t)$ is observed. 
For longer times we observe that detection has a second peak, pinpointing that after the epidemic peak the infectious population falls back under the threshold $\theta$. Finally, Fig.~\ref{fig2}.e-f shows what happens when $\theta=0.07$, but a finite fraction $l_0=0.1$ of the susceptible population is under lockdown. In this case, the pool of susceptible individuals available to be infected is smaller and, as a result, the capacity threshold $\theta$ is more effective in stopping the disease spreading, leading to a better mitigation of the outbreak and, consequently, to a smaller attack rate.

\begin{figure}[t!]
\includegraphics[width=1\linewidth]{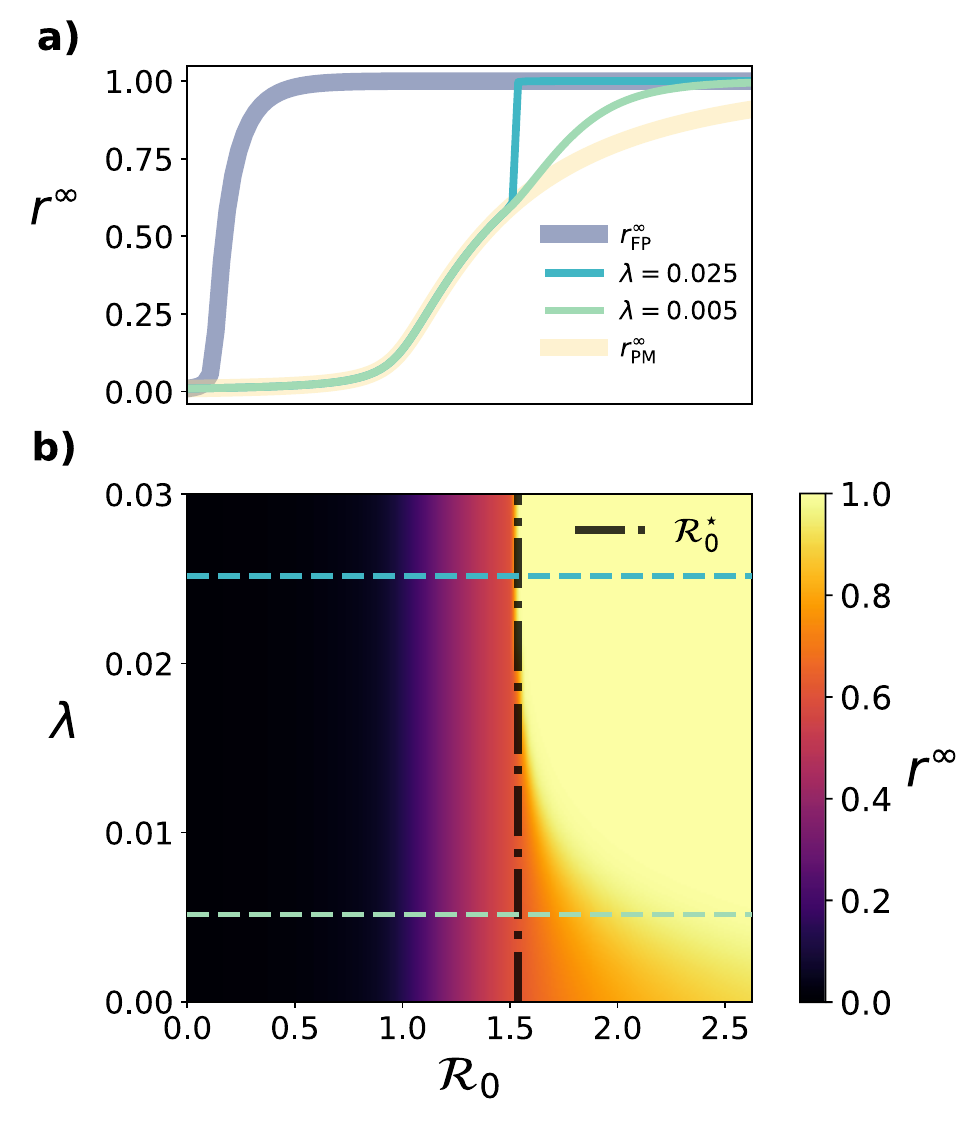}% Here is how to import EPS art
\caption{In panel (a) we show the $SLIDR$ attack rate $r^\infty$ for two values of $\lambda$ in the absence of lockdown ($l_0=0$). The lower value of $\lambda$ yields to a second order collapse transition, while the higher displays a first order transition. The gray curve corresponds to a perfect mitigation case and the red to the free propagation dynamics. In panel (b) the complete phase diagram is shown. The horizontal dashed lines indicate the values of $\lambda$ corresponding to the parameters used to represent curves in panel (a). The vertical black dashed line indicates the critical value $\mathcal{R}_0^{\star}$ computed according to Eq.~(\ref{eq:solution}). In both panels simulations are performed for the range of infectivity values $\beta\in[0,0.75]$, assuming a baseline detection rate $g_0=1$, a capacity threshold $\theta=0.07$, a population size $N=700$ with average connectivity $\langle k\rangle=5$, a recovery rate $\mu=1/7$ and a transition rate from $D$ to $R$ regulated by $\gamma=3/20$.}
\label{fig3} 
\end{figure}

The overall effect of detection and its limited capacity can be analyzed by computing the epidemic diagram, i.e. the impact of the contagion wave, measured by the value of $r^{\infty}$, as a function of the basic reproduction number of the pathogen, ${\cal R}_{0}$, whose expression for the $SLIDR$ model considering an initially fully susceptible population is given by: 
\begin{equation}
    \mathcal{R}_0=\frac{ \beta(\langle k\rangle-1)}{g_0+\mu}\;.
    \label{eq:r0}
\end{equation}
In Fig.~\ref{fig3}.a, we show (thin curves) the epidemic diagrams, $r^{\infty}({\mathcal R}_0)$, in the case $l_0=0$ and a force of detection characterized by a baseline rate $g_0=1$, a capacity threshold $\theta=0.07$, and two different values of $\lambda$. From these diagrams, we observe that there exist two transition points. First, the well-known epidemic threshold at $\mathcal{R}_0=\mathcal{R}_0^c=1$, pinpointing that beyond this point the infective power $\beta(\langle k\rangle-1)$ is larger than the effective recovery rate $(g_0+\mu)$. In addition to the epidemic threshold, a second transition point appears at ${\mathcal R}_{0}^{\star}>1$ which corresponds to the collapse transition. 

To better illustrate the collapse transition point, we also show the epidemic diagrams (thick curves) corresponding to free propagation, $r^{\infty}_{\text{FP}}(\mathcal{R}_0)$, i.e. in the absence of the detection policies, and that corresponding to perfect mitigation,  $r^{\infty}_{\text{PM}}(\mathcal{R}_0)$, which is computed assuming the availability of unlimited resources ($\theta=1$). With these two phase diagrams as limiting cases, it is clear that the collapse point ${\mathcal R}_0^{\star}$ corresponds to the minimum reproduction number that a pathogen needs to jeopardize the detection capacity and decrease the mitigation effects of the early removal of infectious agents. Beyond this point, the epidemic diagram separates from the one corresponding to unlimited detection resources and eventually reaches that corresponding to null detection at a value ${\mathcal R}_0>{\mathcal R}_0^{\star}$ for which the mitigation power of detection is completely suppressed. 

It is also important to discuss the role played by the decay rate $\lambda$. Although $\lambda$ does not intervene in the precise value of ${\mathcal R}_0^{\star}$, it controls the transition between perfect-mitigation and the free-propagation regimes. In particular, the larger the values of $\lambda$, the more abrupt the collapse transition becomes. For large values of $\lambda$, the explosive nature of the collapse transitions implies that once the system reaches the collapse transition $\mathcal{R}_0^{\star}$, the attack rate will be equivalent to the one observed in an uncontrolled scenario. The increase of the sharpness of the collapse transition is clear from Fig.~\ref{fig3}.b where $r^{\infty}(\mathcal{R}_0)$ is reported for a continuous range of $\lambda$ values.

To better characterize the collapse transition shown in  Fig.~\ref{fig3} we have introduced the mitigation efficiency $\eta$, which quantifies the global impact of the detection collapse. The mitigation effectiveness $\eta$, compares the actual epidemic diagram $r^{\infty}({\mathcal{R}}_0)$ for a given detection force ($g_0$, $\theta$, $\lambda$) with that 
obtained with the same detection rate $g_0$ and unlimited resources ($\theta=1$) and is defined as:
\begin{equation}
\eta=\frac{\int_{\mathcal{R}_0}\left(r^{\infty}_{\text{FP}}(\mathcal{R}_0)-r^{\infty}(\mathcal{R}_0)\right)d\mathcal{R}_0}{\int_{\mathcal{R}_0}\left(r^{\infty}_{\text{FP}}(\mathcal{R}_0)-r^{\infty}_{\text{PM}}(\mathcal{R}_0)\right)d\mathcal{R}_0}\;.
\label{eq:mitigacion}
\end{equation}
It takes values in the range $[0,1]$, with $\eta=0$ when the mitigation effect is null, and $\eta=1$ when the mitigation attained is the best possible one. 

\begin{figure}[t!]
\includegraphics[width=1\linewidth]{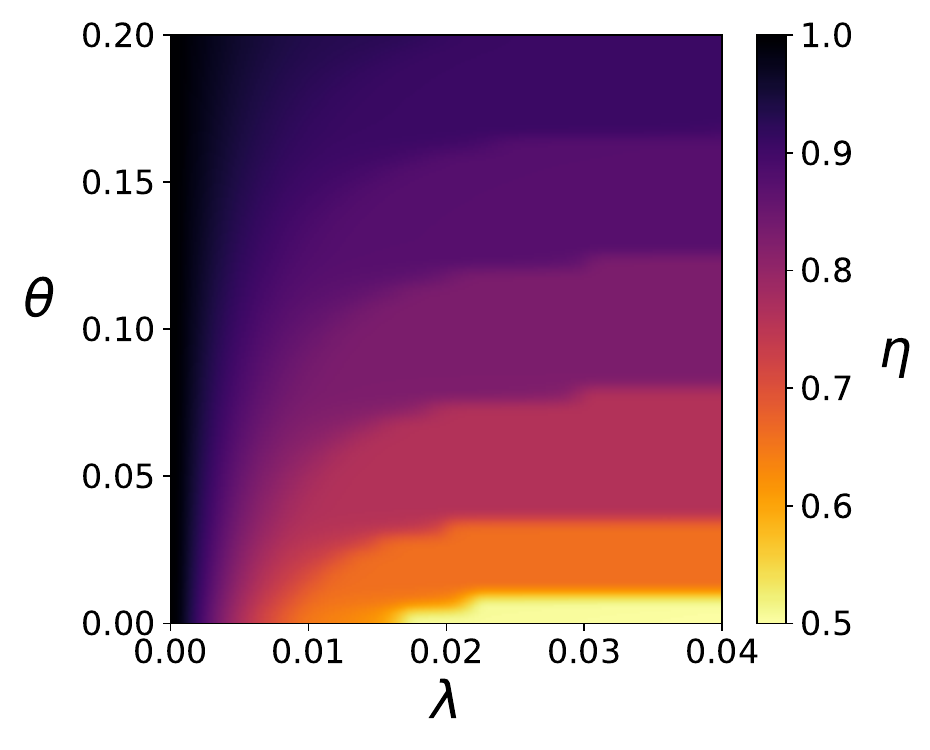}% Here is how to import EPS art
\caption{The mitigation effectiveness $\eta$ (color code) is reported as a function of the capacity threshold $\theta$ and of the decay rate $\lambda$. Simulations are performed for a large range of infectivity values with no lockdown polices implemented ($l_0=0$). We assume a baseline detection rate $g_0=1$, a population size $N=700$ with average connectivity $\langle k\rangle=5$, a recovery rate $\mu=1/7$ and a transition rate from $D$ to $R$ regulated by $\gamma=3/20$.}
\label{fig4} 
\end{figure}

Fig.~\ref{fig4} shows the mitigation effectiveness as a function of $\theta$ and $\lambda$. The latter parameter becomes more relevant when little resources are available, yielding a great variance in the parameter $\eta$ when considering low $\theta$ values. However, as the availability of resources increases, the transition is delayed and the nature of the transition is less relevant. This is because for high basic reproduction number diseases, the free-propagation and the perfect-mitigation curves are quite close, as shown in Fig.~\ref{fig3}.a. 

\section{The collapse threshold ${\cal R}_0^{\star}$}
\label{sec:tres}
Once characterized numerically the existence of a collapse transition leading to the failure of the epidemic containment through detection of infectious agents, we now proceed to derive analytically the precise value of ${\cal{R}}_{0}^{\star}$. This analysis will shed light into the dependence of ${\cal{R}}_{0}^{\star}$ on the epidemiological parameters characterizing the spread of the pathogen. 

As mentioned earlier, the collapse threshold $\mathcal{R}_0^{\star}$ is the minimum value of the basic reproduction number to keep the epidemic curve always below the capacity threshold $\theta$. Therefore, $\mathcal{R}_0^{\star}$ can be determined as the value of $\mathcal{R}_0$ that gives rise to $i_{\text{max}}=\theta$. Although our $SLIDR$ model includes two additional compartments $L$ and $D$, we can effectively treat it as an $SI\bar R$ model with a $\bar {R}$ compartment which aggregates the populations $L$,$D$ and $R$ of the original $SLIDR$ formulation. Since
we assume that the fraction of population in $L$ is contained in the new compartment $\bar{R}$, we have $\bar{r}_0=l_0$ and $s_0=1-l_0-i_0$. Secondly, using our time-continuous approach, we obtain the transition rate from the $I$ compartment to the new ${\bar{R}}$ compartment by adding the detection rate $g_0$ and the recovery rate $\mu$. This way, we can transfer the outgoing flow from $I$ to the effective compartment $\bar{R}$, neglecting the internal dynamics $D\rightarrow R$ within the new effective compartment. After this reformulation we have :
\begin{eqnarray}
%    \frac{ds}{dt}=-(\langle k\rangle-1)\beta si
    \dot s&=&-\left(\langle k\rangle-1\right)\beta si\;,
    \label{eq:s2}\\
    \dot i&=&\left(\langle k\rangle-1\right)\beta si-\left(\mu+g_0\right) i\;,
    \label{eq:i2}\\
    \dot{\bar r}&=&(\mu+g_0) i\;,
    \label{eq:r_ef}
\end{eqnarray}
with initial conditions: $s_0=1-l_0-i_0$, $i_0\ll 1$, and $\bar{r}_0=l_0$. Note that in the formulation of this effective model we have fixed the detection rate to $g_0$ as we are interested in the maximum possible value of $\mathcal{R}_0^{\star}$ not triggering the decrease of the detection rate during the epidemic evolution.

To obtain an analytical expression of $i_{\text{max}}$ we proceed as usual in the SIR model by dividing Eq.~(\ref{eq:s2}) and Eq.~(\ref{eq:r_ef}), yielding:
\begin{equation}
\frac{ds}{d\bar r}=-\frac{(\langle k\rangle -1)\beta}{(g_0+\mu)}s=-\mathcal{R}_0 s\;,
\end{equation}
which can be integrated with the initial conditions above to obtain the evolution of infectious agents as a function of the fraction of susceptible ones:
\begin{equation}
i(t)=1-l_0-s(t)+\frac{1}{\mathcal{R}_0}\log\left(\frac{s(t)}{1-l_0}\right)\;,
\label{eq:i(s)}
\end{equation}
where we have set $\bar{r}_0=l_0$ and made the approximation $s_0\simeq 1-l_0$, considering that, at the beginning of an epidemic, the number of infectious agents is small, $i_0\ll1$.  

The implicit expression $i(s)$ in Eq.~(\ref{eq:i(s)}) allows us to determine the density of susceptibles when the outbreak reaches its peak. Setting $\frac{di}{ds}=0$ we obtain $s(i_{\text{max}})=\mathcal{R}_0^{-1}$ and, by inserting this value into Eq.~(\ref{eq:i(s)}), we finally obtain:
\begin{equation}
i_{\text{max}}=1-l_0-\frac{1}{\mathcal{R}_0}\left[1+\log\left({\mathcal{R}_0(1-l_0)}\right)\right]\;,
\label{eq:imaxgeneral}
\end{equation}

\begin{figure}[t!]
\includegraphics[width=0.95\linewidth]{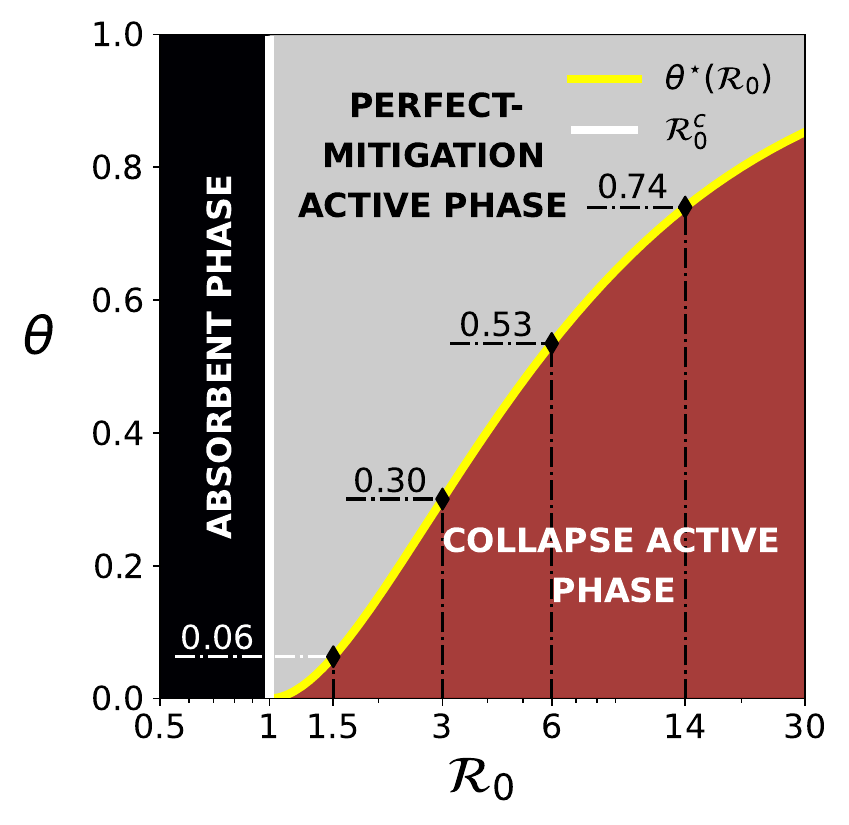}% Here is how to import EPS art
\caption{Phase diagram of the $SLIDR$ model in absence of lockdown ($l_0=0$). $\mathcal{R}_0^c=1$ is the epidemic threshold that separates the absorbent phase from the active epidemic region. Within the active phase, the model allows the computation of the minimum amount of resources $\theta^{\star}$ that we need to avoid collapse, given by (\ref{eq:theta_star1}), as a function of the basic reproduction number $R_0$. Thus, $\mathcal{R}_0^{\star}$ is the critical point related to the transition between that Perfect-Mitigation Active Phase and the Active Collapse Phase. Some $\mathcal{R}_0$ values have been indicated as reference, with their associated $\theta^{\star}$.}
\label{fig:fases_t} 
\end{figure}

Once derived the expression for $i_{\text{max}}$ we set in Eq.~(\ref{eq:imaxgeneral}) the condition fulfilled at $\mathcal{R}_0^{\star}$, i.e. $\theta=i_{\text{max}}$, and after some algebra we obtain the implicit relation for ${\mathcal{R}}_0^{*}$, given the values of $\theta$ and $l_0$:
\begin{equation}
{\mathcal{R}_0^{\star}}(1-l_0)e^{{\mathcal{R}}_0^{\star}(1-l_0)\left[\frac{\theta}{(1-l_0)}-1\right]}=\frac{1}{e}\;.
\label{eq:condition}
\end{equation}
To solve the former relation for ${\mathcal{R}}_0^{\star}$ we perform the following change of variables $x=(\theta/(1-l_0)-1)e^{-1}$ and $y={\mathcal{R}}_0^{\star}(1-l_0)\left[\theta/(1-l_0)-1\right]$. With this new variables the former expression reads:
\begin{equation}
ye^{y}=x\;,
\label{eq:condition2}
\end{equation}
which is a well-known transcendental equation whose solution is the Lambert function, $y=W(x)$.

In order to assess the validity of the expression in Eq.~(\ref{eq:condition2}) to derive ${\mathcal{R}}_0^{*}$ we recall some of the properties of the Lambert function. First, we note that the Lambert $W(x)$ function is only real valuated in the case $x\geq-\frac{1}{e}$. Considering the expression of $x$, it is easy to derive that the former condition demands that $\theta\geq 0$, which is automatically satisfied. Additionally, for $x<0$ ($\theta+l_0<1$), the Lambert function possesses two branches, namely $y=W_0(x)$ and $y=W_{-1}(x)$. This particular range of $x$ is of interest, as it corresponds to situations in which both the capacity threshold and the fraction of locked individuals are small enough for the collapse transition to show up. 

After comparing the numerical solution of ${\mathcal{R}}_0^{\star}$ with the analytical values obtained from the two branches of the Lambert function predictions, we found that the correct behavior is captured by the $y=W_{-1}(x)$ branch. Thus, the expression of the collapse threshold can be finally written as:
\begin{equation}
{\mathcal{R}}_0^{\star}(\theta,l_0)=\frac{1}{\theta-(1-l_0)}W_{-1}\left(\frac{\theta-(1-l_0)}{e(1-l_0)}\right)\;.
\label{eq:solution}
\end{equation}

In Fig.~\ref{fig3}.b we show (see vertical dashed line) that the analytical expression in Eq.~(\ref{eq:solution}) works fairly well in reproducing the precise value of $\mathcal{R}_0$ at which the numerical curve $r^{\infty}(\mathcal{R}_0)$ detaches from the one corresponding to perfect mitigation, $r_{\text{PM}}^{\infty}(\mathcal{R}_0)$, and starts approaching that corresponding to free propagation, $r_{\text{FP}}^{\infty}(\mathcal{R}_0)$.
\begin{figure*}[t!]
\includegraphics[width=1\linewidth]{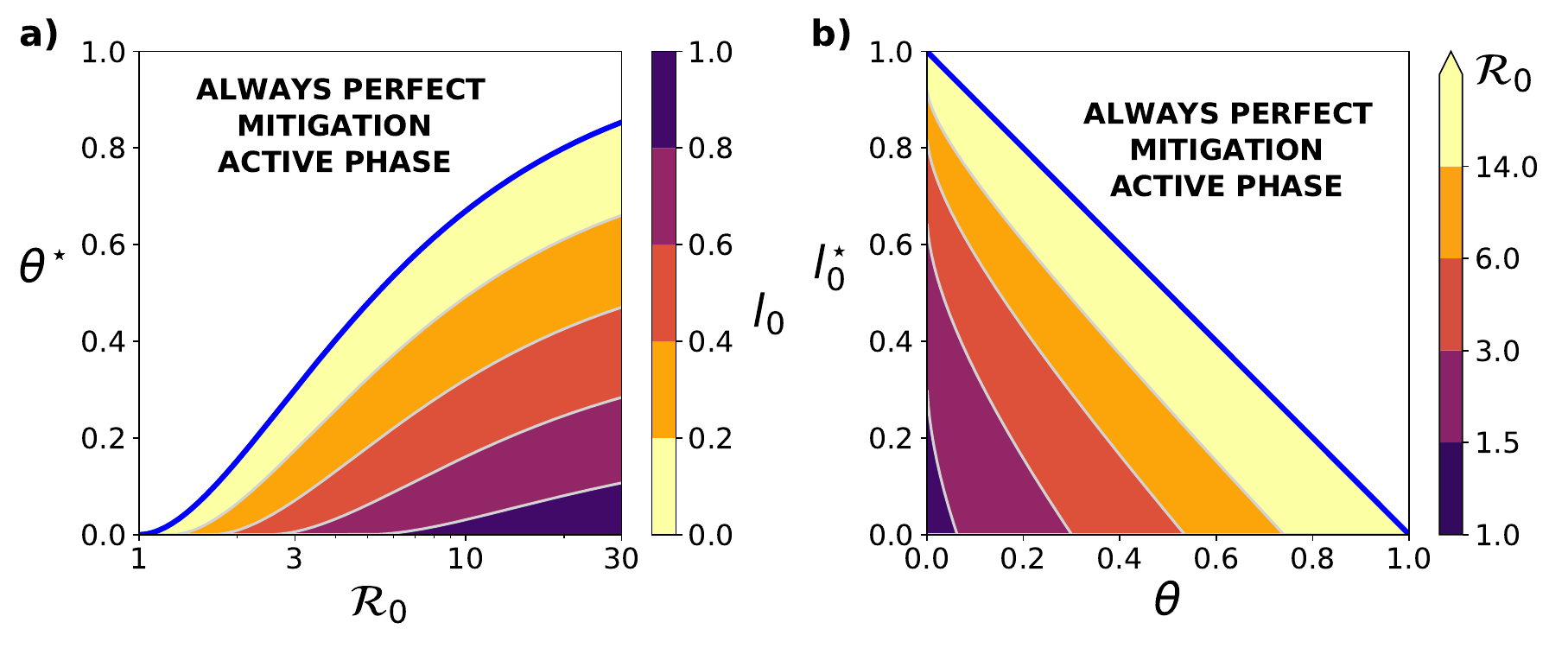}% Here is how to import EPS art
\caption{
In panel (a) we show the critical value $\theta^{\star}(\mathcal{R}_0,l_0)$ according to Eq.~(\ref{eq:theta_star1}). The blue line separates the region of values where the collapse transition can occur from the region where the transition never exists because there is always perfect mitigation. In panel (b) the critical value $l_0^{\star}(\mathcal{R}_0,\theta)$ is drawn according to Eq.~(\ref{eq:lostar}). The blue line indicates the boundary $\theta+l_0=1$ and separates the region where $R_{0}^{\star}$ is finite from the region where the transition never exists. Note that the contour lines correspond to non equally spaced $\mathcal{R}_0$ values.}
\label{fig6} 
\end{figure*}

We continue by addressing the question about what is the minimum amount of resources $\theta$ needed to avoid collapse given that the spreading pathogen is characterized by $\mathcal{R}_0$. This threshold value for $\theta$, hereafter called $\theta^{*}$, can be straightforwardly calculated by just imposing $\theta=i_{\text{max}}(R_0)$ in Eq.~(\ref{eq:imaxgeneral}) which yields:
\begin{equation}
\theta^{\star}(\mathcal{R}_0,l_0)=(1-l_0)\left\{1-\frac{1}{\mathcal{R}_0(1-l_0)}\left[1+\log\left({\mathcal{R}_0}(1-l_0)\right)\right]\right\}\;.
\label{eq:theta_star1}
\end{equation}
Note that the expression (\ref{eq:theta_star1}) can be obtained as the inverse function of the equation (\ref{eq:solution}) in the parameter range ($\theta+l_0<1$) prescribed above.

To complete our analytical derivations, we will now focus on determining the minimum lockdown fraction, denoted by $l_0$, required to maintain perfect mitigation when facing a spreading pathogen characterized by the basic reproduction number $\mathcal{R}_0$, assuming a fixed resource capacity $\theta$.

We obtain the threshold value $l_0^{\star}$ by imposing $\theta=i_{\text{max}}(R_0)$ in Eq.~(\ref{eq:imaxgeneral}). By constructing the Lambert function, we obtain the following expression for the threshold value:
\begin{equation}
l_0^{\star}(\mathcal{R}_0,\theta)=1+\frac{1}{\mathcal{R}_0}W_{-1}\left(-e^{-\theta\mathcal{R}_0-1}\right)\;,
\label{eq:lostar}
\end{equation}
where $W_{-1}$ denotes the branch of the Lambert function. It is worth noting that the validity limits of the Lambert function are always satisfied, since $\theta\mathcal{R}_0\geq0$ by definition of the parameters. Additionally, the branch $W_{-1}$ is well-defined, as the condition $e^{-\theta\mathcal{R}_0-1}>0$ is always fulfilled.

\section{Epidemic phase diagrams}

The analytical results derived in section \ref{sec:tres} allow us to analyze the epidemic phase diagram of the system as a function of the relevant parameters: the infectiousness of the pathogen, $\mathcal{R}_0$, the maximum detection capacity of the system, $\theta$, and the fraction of the population out in shelter, $l_0$. In the first case when $l_0=0$ so that no lockdown is imposed, as shown in Fig.~\ref{fig:fases_t}, the phase diagram can be plotted as a function of $\mathcal{R}_0$ and $\theta$. The diagram exhibits three possible phases: the disease-free phase when $\mathcal{R}_0<1$, the perfect-mitigation phase when $\mathcal{R}_0>1$ and $\theta>\theta^{*}({\mathcal{R}_0})$, and the collapse active phase provided $\mathcal{R}_0>1$ and $\theta<\theta^{*}({\mathcal{R}_0})$. The two former phases are separated by the curve provided by Eq.~(\ref{eq:solution}) (or alternatively Eq.~(\ref{eq:theta_star1})).

The phase diagram in Fig.~\ref{fig:fases_t} shows that for pathogens with $\mathcal{R}_0>3$ the maximum detection capacity should be $\theta^*>0.3$, pointing out that the sole active detection demands an extraordinary amount of resources to avoid the collapse phase. Thus, in the following we explore how combining active detection and lockdown can suffice to achieve the mitigation of the outbreak. From Eq.~(\ref{eq:theta_star1}) it is clear that when lockdown enters into the game ($l_0>0$) two beneficial effects show up. First, the basic reproductive number turns into an lower effective one $\bar{\mathcal{R}}_0={\mathcal{R}}_0(1-l_0)$ and, second, the maximum capacity detection is effectively increased from $\theta^*$ to $\bar{\theta}^*=\theta^*/(1-l_0)$. These two effects combined allow to increase the area of the perfect mitigation phase in the $({\mathcal{R}}_0,\theta)$-plane as shown in Fig.~\ref{fig6}.a which reports the curves $\theta^{*}(\mathcal{R}_0)$ that separate the perfect mitigation and the active collapse phases for different values of the fraction $l_0$. 

The beneficial effects of combining partial lockdown with detection are also illustrated in Fig.~\ref{fig6}.b. Here we show for different values of $\mathcal{R}_0$ the lockdown fraction, $l_0^*$, needed to remain in the perfect mitigation phase considering a fixed maximum detection capacity $\theta$.

\section{Conclusions}

The implementation of reliable detection systems is key to ensure that policies, such as contact tracing and isolation of infectious individuals have the desired impact on outbreak control. In this study, we have introduced and studied a compartmental model in which detection resources are limited. Using a mean field approach, we have characterized the different dynamic regimes of the system as a function of its parameters. 

The most relevant result of this work is the observation of two transitions as a function of the basic reproductive number: the epidemic ($\mathcal{R}_0=1$) and the collapse ($\mathcal{R}_0=\mathcal{R}_0^{\star}$) transitions. In the latter transition the health system is unable to meet the demand for detection (for $\mathcal{R}_0>\mathcal{R}_0^{\star}$) and we move from a controlled regime, where detection drives the mitigation of the epidemic outbreak, to a regime in which the pathogen spreads freely. 

The existence of a collapse transition has motivated the analysis of a combined implementation of detection and lockdowns~\cite{Peak2020Individual}. We have observed that the combination of the two strategies can help to avoid the collapse point, specially for those pathogens with a large $\mathcal{R}_0$. Besides, our results show how, for certain values of the decay constant $\lambda$, the nature of the collapse transition turns out to be explosive. This result is striking because it means that, above the collapse transition, the attack rate can be the same as in the unmitigated dynamics. The way on how this explosive behavior shows up is also remarkable since it arises in the active phase, i.e. well beyond the epidemic threshold.  Thus, it stands in contrast to conventional forms of explosivity observed in contagion models \cite{expcontagion,Transicion}, in which deliberate delays in epidemic onset lead to abrupt transitions deviating from the usual smooth ones \cite{review_exptran}.

Finally, it should be noted that the analytical results presented here are based on a mean-field approach and, thus, some limitation are worth mentioning. First our model does not account for all observed features of social connections and pathogen performance. In this context, it would be worthwhile to refine the mechanisms built into $SLIDR$ to be able to incorporate real connectivity patterns given by networks of close contacts. In this area, the inclusion of contact tracing strategies and not only symptomatic detection could be of particular interest. This approach could also be applied to reaction-diffusion processes that simultaneously incorporate mobility flows and contact patterns \cite{cota2021}, paving the way for the identification of optimal distributions of detection resources ~\cite{zhu2021allocating,zhang2022intervention}. In addition, the lockdown that complements detection has been implemented in a stylized way, i.e. starting from the beginning of the epidemic way rather than being applied in subsequent times. For this particular scenario, we have checked that the reported results are robust provided the time elapsed between the start of the epidemic wave and the lockdown is small enough compared to the time associated to the epidemic peak. Overall, our model is able to provide analytical insights as a benchmark for more realistic models that can capture the full range of complexities involved in infectious disease outbreaks.

\section*{Acknowledgements}
 S.L.O and J.G.G. acknowledge financial support from the Departamento de Industria e Innovaci\'on del Gobierno de Arag\'on y Fondo Social Europeo (FENOL group grant E36-23R) and from grant PID2020-113582GB-I00 funded by MCIN/AEI/10.13039/501100011033.

\bibliography{refs}

\end{document}